\documentclass[a4paper,fleqn,usenatbib]{mnras}

\usepackage{newtxtext,newtxmath}
\usepackage[T1]{fontenc}
\usepackage{ae,aecompl}

\usepackage{float}
\usepackage{amsmath}
\usepackage{amssymb, soul} %, graphics, setspace, color, soul}
\usepackage{multirow}
\usepackage{graphicx}
\usepackage{natbib}

\usepackage{hyperref}

\newcommand{\dif}{\mathrm{d}}
\soulregister\cite7 % 针对\cite命令
\soulregister\citep7 % 针对\citep命令
\soulregister\citet7 % 针对\citet命令
\soulregister\ref7 % 针对\ref命令
\soulregister\pageref7 % 针对\pageref命令

\title[Yarkovsky effect \& Eos family ]{Asteroid migration due to the Yarkovsky effect and the distribution of the Eos family}

\author[Y.-B. Xu et al]{
	Yang-Bo Xu,$^{1,2}$
	Li-Yong Zhou,$^{1,2,5}$\thanks{E-mail: zhouly@nju.edu.cn}
	Christoph Lhotka,$^{3}$ 
	Wing-Huen Ip$^{4}$
	\\
	$^{1}$School of Astronomy and Space Science, Nanjing
	University, Nanjing 210046, China\\
	$^{2}$Key Laboratory of Modern
	Astronomy and Astrophysics in Ministry of Education, Nanjing
	University, Nanjing 210046, China\\
	$^{3}$Space Research Institute,
	Austrian Academy of Sciences, 8042 Graz, Austria\\
	$^{4}$Institute of Astronomy, National Central University, Taiwan \\
    $^{5}$Institute of Space Astronomy and Extraterrestrial Exploration (NJU \& CAST), China}

\date{Accepted XXX. Received YYY; in original form ZZZ}

\pubyear{2019}

\begin{document}
\label{firstpage}
\pagerange{\pageref{firstpage}--\pageref{lastpage}}
\maketitle

\begin{abstract}
Based on a linearized model of the Yarkovsky effect, we investigate in this paper the dependence of the semimajor axis drift $\Delta a$ of a celestial body on its size, spinning obliquity, initial orbit and thermal parameters on its surface. 
With appropriate simplification and approximation, we obtain the analytical solutions to the  perturbation equations for the motion of asteroids influenced by the Yarkovsky effect, and they are then verified by numerical simulations of the full equations of motion. These solutions present explicitly the dependencies of $\Delta a$ on the thermal and dynamical parameters of the asteroid. With these analytical formulae for $\Delta a$, we investigate the combined seasonal and diurnal Yarkovsky effects. The critical points where the migration direction reverses are calculated and the consequent selective effects according to the size and rotation state of asteroids are discussed. %Solely the Yarkovsky effect is found to be able to produce some ring structure in the aged circumstellar debris disk. 
Finally, we apply the analytical formulae to calculate the migration of Eos family members. The space distribution of asteroids is well reproduced. Our calculations suggest that statistically the orientations of spin axes of family members satisfy a random-obliquity distribution, and the rotation rate $\omega_{\rm rot}$ of asteroid depends on its size $R$ by $\omega_{\rm rot}\propto R^{-1}$. 
\end{abstract}


%
%________________________________________________________________

\section{Introduction}
The Yarkovsky effect manifests itself in the form of a recoil force due to thermal radiation from anisotropically heated orbiting bodies. For asteroids in the Solar system, the side of the body facing to the Sun is heated through absorption of the solar radiation. After a short period of time, the asteroid re-radiates its thermal energy. However, due to the
rotation and revolution of the asteroid, its hottest side is not aligned with the Sun anymore, so that the thermally induced radiation pressure has a transverse component, which will accelerate or de-accelerate the asteroid orbital motion, and thus have an effect on the semimajor axis on reasonable long time scales.

The effect was discovered by an unknown Polish/Russian engineer, Ivan O. Yarkovsky, around the year 1900. The resulting force is weak and was not noticed for a long time until \"Opik proposed it again to better understand the motion of meteoroids. %Although he just gave some estimates for the magnitude, his idea influenced future researchers.  
\citet{rad52,pet76,bur79} did a lot of detailed studies and derived very useful results. The study on the motion of artificial satellite LAGEOS inspired people to generalize the classical Yarkovsky effect. To explain the unexpected variation of the along-track acceleration of LAGEOS, \citet{rub87} developed a ``LAGEOS-taylored'' technique for computing the thermal force perturbations on a rapidly rotating body. Recognizing the Yarkovsky effect as an important perturbation to the motion of artificial satellite, Rubincam proposed that the Yarkovsky effect consisted of two components, the ``diurnal'' and the ``seasonal'' effect. \cite{rub95} first tried to reconcile the modelling of Yarkovsky effect on LAGEOS and natural celestial  bodies. 
The time needed for reducing the semimajor axis of a basaltic asteroid (60\,m in radius and on a circular low-inclination orbit) by 2\,AU was calculated. In addition, the variation of the eccentricity for low-inclination orbits was discussed and it was found that the Yarkovsky effect could circularize orbits. 

\citet{far98} showed that the diurnal Yarkovsky effect plays an important role in the delivery of meteorites and the dynamics of small bodies in the Solar system. The drift rates of  semimajor axis were calculated for celestial bodies of different sizes that are of three kinds of materials and at three values of initial semimajor axis. Furthermore, \citet{far99} calculated the semimajor axis displacements of asteroids with different radii and for several values of thermal conductivity within the collisional life. Based on the previous model \citep{vok98a}, \citet{vok99} developed an accurate linear model for the Yarkovsky thermal force on spherical asteroids. The diurnal and seasonal components were combined together and additional mixed terms were derived. 
A series of work \citep{vok98a,vok98b,vokfar99,vokbot01} investigated in great detail the Yarkovsky effect, about the non-spherical bodies, the nonlinear theory, the albedo choice, etc. Noting that most aforementioned results were derived for circular orbits, \citet{spi01} did numerical evaluation of the Yarkovsky effect for non-circular orbits, focusing on the relationship between the semimajor axis' drift rate and its initial value.

With the mathematical model of the Yarkovsky effect being completed, the application of it to motion of small celestial bodies began to receive attentions. For example, \citet{bot01} studied the dynamical spreading of Koronis family by the Yarkovsky Effect. \citet{tsi03} demonstrated that the short-lived asteroids in the 7/3 Kirkwood gap were replenished by members of the Koronis and Eos families pushed by the Yarkovsky effect. \citet{spo15} used the so-called V-shapes of asteroid families produced by the Yarkovsky effect to estimate ages of these families. \citet{marz13} studied the influence of Yarkovsky effect on the motion of Earth Trojans, and it was recently found that all possible primordial Earth Trojans should have been driven out of the 1/1 resonance region by this effect \citep{zhou19}. And lately, \citet{chris20} investigated the combined Yarkovsky and YORP effects on the population of Mars Trojans. 

The improving accuracy of observations brings more and more facts that demonstrate the importance of Yarkovsky effect in the dynamics of asteroids in the Solar system. So far, many  mathematical analyses and numerical simulations have been made. However, to find the straightforward relationship between the strength of Yarkovsky effect and the parameters of the involved celestial body, especially the thermal parameters, still deserves a closer investigation. Based on the brilliant work in the literature, in this paper we derive an analytical approximate solution of the semimajor axis drift, so that a more complete knowledge of this thermal effect can be achieved.

By both analytical method and numerical simulations, we present in this paper the dependence of the Yarkovsky effect on the dynamical and thermal parameters. In Section 2, we briefly summarize the theory of Yarkovsky effect for a simplified model of a homogeneous spherical asteroid. Both the seasonal and diurnal effects are given in a unified form. In Section 3, after appropriate approximation and simplification, for the first time we give the analytical formulae for the semimajor axis drift ($\Delta a$), in which the dependences of $\Delta a$ on thermal parameters and dynamical parameters are given explicitly. In Section 4, we calculated numerically $\Delta a$ to verify our analytical results and present the variation of $\Delta a$ with respect to different parameters. In Section 5, the seasonal and diurnal effects, and the direction of the Yarkovsky migration are discussed using the analytical formulae. As an example of the practicability and advantages of our analytical formulae, in Section 6 we apply the analytic solutions to discuss the distribution of semimajor axes of Eos family members. We summarize the paper in Section 7.

%__________________________________________________________________

\section{Theory}

\subsection{Yarkovsky force model} \label{sec:mod}
We just describe briefly the basic model of the Yarkovsky effect for reference in this section. For more details, please refer to e.g. \cite{vok99}.

The Yarkovsky effect is mainly determined by the temperature distribution on the surface of an asteroid, which in turn is determined by its thermal properties. The heat conduction in asteroids can be described by the Fourier equation of temperature $T$:
   \begin{equation} 
      \rho C \frac{\partial T}{\partial t} = K\nabla^2T,
   \label{eq:fourier}
   \end{equation}
where $\nabla^2$ is the Laplace operator, $K, C, \rho$ are the thermal conductivity, specific heat capacity and density of the material, respectively. The boundary condition is provided by the conservation of energy
   \begin{equation} \label{eq:boundary}
      \epsilon\sigma T^4 + K \left(\mathbf{n}\cdot \frac{\partial T}{\partial \mathbf{r}}\right) = \alpha \mathcal{E}.
   \end{equation}
The first term on the left-hand side accounts for the energy thermally reradiated by the asteroid, and the second term gives the energy conducted to deeper layers of the body. The right-hand side gives the radiation energy entering the unit surface area per unit time. As for the notations, $\epsilon$ denotes the emissivity, $\sigma$ the Stefan-Boltzmann constant, $\mathbf{n}$ the unit vector normal to the surface, $\alpha$ the absorption coefficient and $\mathcal{E}$ the external radiation flux. 

With equations~\eqref{eq:fourier} and \eqref{eq:boundary}, a distribution of temperature $T$ throughout the asteroid at any time $t$ can be derived. This is done in a rotating, body-fixed reference frame, with the $Z$-axis coinciding with the spin vector. At the initial time $t_0$, the $X$-axis of this system points toward the radiation source. The $Y$-axis completes the right hand side coordinate system.

Due to the existence of the fourth-power emission law in the first term of equation~\eqref{eq:boundary}, a general solution is fairly complicated. It is reasonable to assume that the temperature throughout the body does not differ too much from the average value (i.e. $\Delta T\ll \bar{T}$). Then the emission term can be linearised as $T^4\approx \bar{T}^4+4\bar{T}^3\Delta T$. The asteroid is generally supposed to be a spherical body so that it can be parametrized by spherical coordinates $(r, \theta, \phi)$. %that will be adopted below.

To simplify the mathematical formulation of the problem, some auxiliary variables will be introduced first. The temperature $T$ will be normalized by $T_\star$: $\epsilon\sigma T^4_\star=\alpha\mathcal{E}_\star$, where $\mathcal{E}_\star$ denotes the solar radiation flux at the mean distance from the Sun along the orbit. Similarly, the radial coordinate $r$ measured from the centre of the body to its surface (at $r=R$) is to be scaled by the
penetration depth $l_s$ of the seasonal thermal wave ($l_s=\sqrt{K/\rho C\omega_{\rm rev}}$, where $\omega_{\rm rev}$ is orbital revolution frequency of the asteroid),  $r'_s=r/l_s$. Finally, the time $t$ will be represented by a complex quantity $\zeta=\exp(\mathrm{i}\lambda)$, where $\lambda=\omega_{\rm rev}(t-t_0)$. With these auxiliary variables, equations~\eqref{eq:fourier} and \eqref{eq:boundary} turn to be
   \begin{equation} \label{eq:fourier1}
      \mathrm{i}\zeta\frac{\partial}{\partial\zeta}\Delta T' = \frac{1}{{r'_s}^2}\left[\frac{\partial}{\partial r'_s}\left({r'_s}^2\frac{\partial}{\partial r'_s}\right)+\Lambda\right]\Delta T',
   \end{equation}
   \begin{equation} \label{eq:boundary1}
      \sqrt{2}\Delta T'+\Theta_s\left(\frac{\partial\Delta T'}{\partial r'_s}\right)_{R'_s} = \Delta\mathcal{E}',
   \end{equation}
with
\begin{equation}
\Lambda=\frac{1}{\sin\theta}\left[\frac{\partial}{\partial\theta}\left(\sin\theta\frac{\partial}{\partial\theta}\right)+\frac{1}{\sin\theta}\frac{\partial^2}{\partial\phi^2}\right],
\end{equation}
\begin{equation} \label{eq:theta}
 \Theta_s=\frac{\sqrt{\rho CK\omega_{\rm rev}}}{\epsilon\sigma T^3_\star}.
\end{equation}
Using spherical functions to expand $\Delta \mathcal{E}'$ and $\Delta T'$, a general solution of equations~\eqref{eq:fourier1} and \eqref{eq:boundary1} for the dipole part can be obtained. And using Lambert's law, the recoil force from the asteroid's scattering and thermal emission can be obtained:
   \begin{equation} \label{eq:force}
      \mathbf{f} = -\frac{2\sqrt{2}}{3\pi}\alpha\Phi\int \Delta T'(R'_s; \theta, \phi; \zeta)\mathbf{n} \, \dif\Omega.
   \end{equation}
In this formula, $\Phi=\mathcal{E}_\star\pi R^2/mc$ is the usual radiation force factor, with $m$ and $c$ being the asteroid's mass and speed of light, respectively. Then, the thermal force components can be calculated as follows.
   \begin{eqnarray} \label{eq:forcecom1}
   \begin{aligned}
         f_X+\mathrm{i}f_Y = & -\frac{4\alpha\Phi}{9(1+\chi)}\Big[ \sin^2\frac{\gamma}{2}E_{R'_+}\exp(-\mathrm{i}\delta_{R'_+})\zeta^{-1}\\
         & \hspace{1.7cm} +\cos^2\frac{\gamma}{2}E_{R'_-}\exp(-\mathrm{i}\delta_{R'_-})\zeta\Big]\zeta^{-\frac{\omega_{\rm rot}}{\omega_{\rm rev}}},          \\
      f_Z = &  \frac{4\alpha\Phi}{9(1+\chi)}\sin\gamma       E_{R'_s}\sin(\lambda+\delta_{R'_s}).
   \end{aligned}
   \end{eqnarray}
Here, $\gamma$ is the obliquity of the spin axis with respect to the norm of orbital plane, $\omega_{\rm rot}$ is the spin (rotation) rate, 
\begin{equation} R'_\pm= R'_d\sqrt{1\pm\frac{\omega_{\rm rev}}{\omega_{\rm rot}}}, \hspace{0.1cm} R'_d= \frac{R}{l_d}, \hspace{0.1cm} l_d= l_s\sqrt{\frac{\omega_{\rm rev}}{\omega_{\rm rot}}}, \hspace{0.1 cm} \chi=\frac{\Theta_s}{\sqrt{2}R'_s},
\end{equation}
and the amplitude $E_{R'}$ and phase $\delta_{R'}$ defined as \citep{vok98a}
   \begin{equation} \label{eq:xdef}
      E_{R'}\exp(\mathrm{i}\delta_{R'}) = \frac{A(x)+\mathrm{i}B(x)}{C(x)+\mathrm{i}D(x)},
   \end{equation}
with $x = \sqrt{2}R'$ and the auxiliary functions $A(x), B(x), C(x), D(x)$
   \begin{eqnarray} \label{eq:auxabcd}
   \begin{aligned}
      A(x) = & -(x+2)-\mathrm{e}^x\left[(x-2)\cos x-x\sin  x\right],\\
      B(x) = & -x-\mathrm{e}^x\left[x\cos x+(x-2)\sin x\right],\\
         C(x) = & A(x)+\frac{\chi}{1+\chi}\\
         & \times \left\{3(x+2)+\mathrm{e}^x\left[3(x-2)\cos x+x(x-3)\sin x\right]\right\}, \\
         D(x) = & B(x)+\frac{\chi}{1+\chi}\\
         &\times \left\{x(x+3)-\mathrm{e}^x\left[x(x-3)\cos x-3(x-2)\sin x\right]\right\}.
   \end{aligned}
   \end{eqnarray}
Note in equation~\eqref{eq:forcecom1} both the seasonal and diurnal effects are given in a unified form \citep{vok99}, thus a radial coordinate normalization by the penetration depth $l_d$ of the diurnal wave has been introduced, and the subscript ``$d$'' indicates such diurnal-type variables. 

To derive the perturbation equations for the asteroid motion, the force components need to be transformed to the body-centred frame with the axes $\vec{r}, \vec{t}, \vec{n}$ being radial, transverse, and normal to the orbital plane, respectively. For most of the Solar system bodies, their spin periods are much shorter than their revolution periods (hours versus years), so that $R'_\pm\approx R'_d$. The coordinate transform can be easily performed by the rotation matrix $R_Z(\lambda+\pi)R_X(-\gamma)R_Z(-\theta)(f_X,f_Y,f_Z) = (f_r,f_t,f_n)$, and the components of force in the new coordinate frame turn out to be:
   \begin{eqnarray} \label{eq:force2}
   \begin{aligned}
     f_r = & \frac{4\alpha\Phi}{9(1+\chi)}\Bigg(E_{R'_s}\sin\left(\delta_{R'_s}+\lambda\right)\sin\lambda\sin^2\gamma \\
     & + E_{R'_d}\cos\delta_{R'_d}\left(\cos^2\lambda+\sin^2\lambda\cos^2\gamma\right)\Bigg), \\ 
     f_t = & \frac{4\alpha\Phi}{9(1+\chi)}\Bigg(E_{R'_s}\sin\left(\delta_{R'_s}+\lambda\right)\cos\lambda\sin^2\gamma \\
     & -E_{R'_d}\left(\cos\delta_{R'_d}\sin\lambda\cos\lambda\sin^2\gamma+\sin\delta_{R'_d}\cos\gamma\right)\Bigg), \\
     f_n = & \frac{4\alpha\Phi}{9(1+\chi)}\Bigg(E_{R'_s}\sin\left(\delta_{R'_s}+\lambda\right)\sin\gamma\cos\gamma \\
     & -E_{R'_d}\left(\cos\delta_{R'_d}\sin\lambda\sin\gamma\cos\gamma-\sin\delta_{R'_d}\cos\lambda\sin\gamma\right)\Bigg).
   \end{aligned}
   \end{eqnarray}
We notice, that the aforementioned rotation matrix is valid under the assumption that the orbital plane of the asteroid around the Sun only changes slowly.

With the recoil force given above, the perturbation equations of the asteroid can be derived. In this paper, we will focus on the semimajor axis ($a$) drifting due to the Yarkovsky effect.

\subsection{Perturbation equations}  \label{sec:pert}
Assume an asteroid that is initially on a near-circular orbit with $e\ll 1$.  Since $\dif e/\dif t\propto e$ \citep{vok98a}, the eccentricity will not increase significantly but be always close to zero. Substituting the force $(f_r,f_t,f_n)$ in equation~\eqref{eq:force2} into Gauss perturbation equations \citep[see e.g.][]{mur99} and neglecting
$\mathcal{O}(e)$, we get the equations for the semimajor axis averaged over one revolution:
   \begin{eqnarray} \label{eq:aeq}
      \begin{aligned}
        & \left(\frac{\dif a}{\dif t}\right)_s = \frac{4\alpha\Phi}{9\omega_{\rm rev} (1+\chi)}E_{R'_s}\sin\delta_{R'_s}\sin^2\gamma, \\
        & \left(\frac{\dif a}{\dif t}\right)_d = -\frac{8\alpha\Phi}{9\omega_{\rm rev}(1+\chi)}E_{R'_d}\sin\delta_{R'_d}\cos\gamma.
      \end{aligned}
   \end{eqnarray}
The subscripts $s, d$ represent the seasonal and diurnal effect, respectively. Equation~\eqref{eq:aeq} is consistent with the results of \citet{vok99}. It is worth noting that in these equations $\omega_{\rm rev}, \Phi$ and $\frac{E_{R'}\sin\delta_{R'}}{1+\chi}$ are all functions of semimajor axis, being related to $a$ through mean motion, mean solar radiation flux $\mathcal{E}_{\star}$ and mean temperature $T_{\star}$, respectively. Particularly, $\frac{E_{R'}\sin\delta_{R'}}{1+\chi}$ %defined in equation}~\eqref{eq:xdef} 
can be written in an equivalent form \citep[see e.g.][]{vok98a,bot06}, e.g. for diurnal effect, as
\begin{equation}\label{eq:relax}
F(R'_d,\Theta_d)=\frac{\kappa_3\Theta_d}{1+\kappa_1\Theta_d+\kappa_2\Theta_d^2},
\end{equation}
where coefficients $\kappa_{1,2,3}$ are functions of $R'_d$, and the thermal parameter
\begin{equation}
 \Theta_d=\frac{\sqrt{\rho CK\omega_{\rm rot}}}{\epsilon\sigma T^3_\star}=\Theta_s\sqrt{\frac{\omega_{\rm rot}}{\omega_{\rm rev}}}  
\end{equation} 
(see equation~\eqref{eq:theta} for definition of $\Theta_s$) is a measure of the relaxation between the absorption and re-irradiation of the energy for diurnal effect. Obviously, $F(R'_d,\Theta_d)\propto \Theta_d$ when $\Theta_d\ll 1$. Substitute this relation into equation~\eqref{eq:aeq} and a little algebra yields $(\dif a/\dif t)_d\propto a$. If $\Theta_d\gg 1$, then $F(R'_d,\Theta_d)\propto \Theta_d^{-1}$ and similar calculation yields $(\dif a/\dif t)_d\propto a^{-2}$. In this paper, we adopt equation~\eqref{eq:xdef} to facilitate the following calculations. 

Even after these simplifications, the general analytic solutions to these perturbation equations cannot be found. But in some specific situations and with appropriate approximations, the analytic solutions could be derived, as we will show below.

\section{Analytical estimation of Yarkovsky migration} \label{sec:est}
Consider the term $\frac{\Phi}{1+\chi}E_{R'}\sin\delta_{R'}$ in 
equation~\eqref{eq:aeq}. %$R'$ represents $R'_s$ or $R'_d$. 
Recall the definitions given in equation~\eqref{eq:xdef}, we may denote $\chi=c_1/x$, $\Phi=c_2/x$, with $x=\sqrt{2}R'$ and
   \begin{eqnarray} \label{eq:c12def}
   \begin{aligned}
      (c_1)_s = & \frac{2\sqrt{2}(\rho KC)^\frac{1}{2}\pi^\frac{3}{4}\mu^\frac{1}{4}a^\frac{3}{4}}{(\alpha L)^\frac{3}{4}(\epsilon\sigma)^\frac{1}{4}},\\
      (c_2)_s = & \frac{3\sqrt{2}LC^\frac{1}{2}\mu^\frac{1}{4}}{16\pi c(\rho K)^\frac{1}{2}a^\frac{11}{4}},\\
      (c_1)_d = & \frac{2\sqrt{2}(\rho KC)^\frac{1}{2}\pi^\frac{3}{4}a^\frac{3}{2}\omega_{\rm rot}^\frac{1}{2}}{(\alpha L)^\frac{3}{4}(\epsilon\sigma)^\frac{1}{4}},\\
      (c_2)_d = & \frac{3\sqrt{2}LC^\frac{1}{2}\omega_{\rm rot}^\frac{1}{2}}{16\pi c(\rho K)^\frac{1}{2}a^2}.\\
   \end{aligned}
   \end{eqnarray}
Here, $L$ is the luminosity of the Sun, and $\mu=G(M_{\odot}+m)$ is the reduced mass of the system. Using the Kepler's third law, $\omega^2_{\rm rev}a^3=\mu$, the parameters for seasonal and diurnal effects can be related by $(c_{1,2})_d = (c_{1,2})_s\sqrt{\omega_{\rm rot}/\omega_{\rm rev}}$. In addition, a little algebraic calculation finds
\begin{equation} \label{eq:c-theta}
(c_1)_d= \left[\frac{\epsilon\sigma T_{\ast}^4}{\alpha L/(4\pi a^2)} \right]^{\frac{3}{4}}\Theta_d.
\end{equation}
Two cases, of $x \ll 1$ (thus $R'\ll 1$) and of $x \gg 1$ (thus $R'\gg 1$), will be discussed separately below.

\subsection{When $R'\ll 1$}
When the asteroid size $R$ is small compared to the thermal penetration depth $l_s$ or $l_d$, i.e. $R'\ll 1$, the term $\frac{\Phi}{1+\chi}E_{R'}\sin\delta_{R'}$ can be expanded around $x=0$. 
Ignoring the third and higher order terms, 
   \begin{equation}
      \frac{\Phi}{1+\chi}E_{R'}\sin\delta_{R'} = -\frac{c_2}{10c_1}x^2.
   \end{equation}
Substitute it into the perturbation equation~\eqref{eq:aeq}, we get the solutions (evolution of $a$ with time $t$ due to seasonal and diurnal effects, respectively)
   \begin{eqnarray} \label{eq:asol}
    \begin{aligned}
     &  a_s = \left(a_0^\frac{9}{2}-\frac{3(\alpha L)^\frac{7}{4}(\epsilon\sigma)^\frac{1}{4} CR^2\sin^2\gamma}{80\pi^\frac{7}{4}K^2c}t\right)^\frac{2}{9},\\
     &  a_d = \left(a_0^3+\frac{(\alpha L)^\frac{7}{4}(\epsilon\sigma)^\frac{1}{4} CR^2\omega_{\rm rot}\cos\gamma} {20\pi^\frac{7}{4} \mu^\frac{1}{2} K^2c}t\right)^{\frac{1}{3}},
    \end{aligned}
   \end{eqnarray}
where $a_0$ is the initial value of semimajor axis. If $a_s$ and $a_d$ do not change too much, the variations of them, $\Delta a_s, \Delta a_d$ can be further simplified using Taylor series expansion, and they are
%to Eqs.\,(\ref{eq:asol})\,\&\,(\ref{eq:isol}).
   \begin{eqnarray} \label{eq:deltar1}
    \begin{aligned}
     & \Delta a_s \approx -\frac{(\alpha L)^\frac{7}{4}(\epsilon\sigma)^\frac{1}{4} CR^2\sin^2\gamma}{120\pi^\frac{7}{4}K^2ca_0^{\frac{7}{2}}}t,\\
     & \Delta a_d \approx \frac{(\alpha L)^\frac{7}{4}(\epsilon\sigma)^\frac{1}{4} CR^2\omega_{\rm rot}\cos\gamma}{60\pi^\frac{7}{4}\mu^\frac{1}{2}K^2ca_0^2}t.
    \end{aligned}
   \end{eqnarray}

\subsection{When $R'\gg 1$}
When the asteroid size is large, $R'\gg 1$, the $\mathrm{e}^{2x}$ terms in $\frac{\Phi}{1+\chi}E_{R'}\sin\delta_{R'}$ are dominant. Ignoring other terms, it becomes
   \begin{equation} \label{eq:cofrb}
   \begin{split}
      \frac{\Phi}{1+\chi}E_{R'}\sin\delta_{R'} = &c_1c_2(2x^2-x^3)\Big[16c_1^2-16c_1(1+c_1)x\\
      &+4(1+4c_1+2c_1^2)x^2-4(1+c_1)^2x^3\\
      &+(2+2c_1+c_1^2)x^4\Big]^{-1} \\
      \approx & -\frac{c_1c_2}{2+2c_1+c_1^2}\cdot x^{-1}.
   \end{split}
   \end{equation}
Considering the relation between $c_1$ and $\Theta$ as shown by equation~\eqref{eq:c-theta}, we know that it is not a coincidence that the coefficient before $x^{-1}$ in this equation has the same form as the right-hand side of equation~\eqref{eq:relax}. Substitute this into equation~\eqref{eq:aeq}, we may get the analytical solutions to these equations. However, the analytical expressions of these solutions are complicated. Here we would skip these cumbersome expressions but rather present explicitly the approximation of $\Delta a_s$ and $\Delta a_d$ calculated from these solutions.
   \begin{eqnarray} \label{eq:deltarg}
    \begin{aligned}
     & \Delta a_s = \left.-\frac{4\alpha
      (c_1)_s(c_2)_s\sin^2\gamma}{9\omega_{\rm rev} \left[2+2(c_1)_s+(c_1)_s^2\right]x_s}t\right|_{a=a_0},\\
     & \Delta a_d = \left.\frac{8\alpha 
      (c_1)_d(c_2)_d\cos\gamma}{9\omega_{\rm rev} \left[2+2(c_1)_d+(c_1)_d^2\right]x_d}t\right|_{a=a_0}.
    \end{aligned}
   \end{eqnarray}
The thermal parameters $K, C, \rho$ are embodied within $c_1$ and $c_2$ in equation~\eqref{eq:deltarg}, which makes the relation between the Yarkovsky effect and these parameters a little indistinct. In fact, this can be simply improved as follows.

\begin{figure}%[htbp]
\centerline{
\includegraphics[width=9.0cm,height=7.50cm,angle=0]{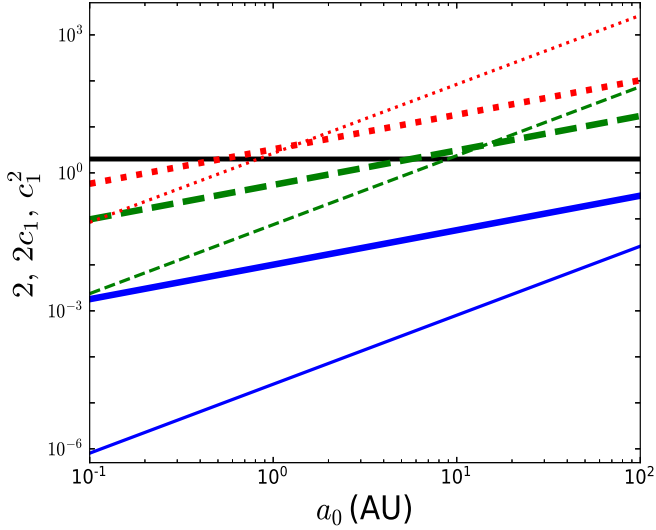}}
\caption{The values of $2(c_1)_s$ (thick lines) and $(c_1)_s^2$ (thin lines) against initial semimajor axis $a_0$ for seasonal effect. The blue solid, green dashed and red dotted lines represent the regolith-covered, basalt and iron-rich asteroids, respectively. The constant value 2 is also plotted for comparison. For the diurnal effect, $(c_1)_d=(c_1)_s\sqrt{\omega_{\rm rot}/\omega_{\rm rev}}$.} 
\label{fig:0}
\end{figure}

The denominator in equation~\eqref{eq:cofrb} includes three terms $2, 2c_1$ and $c^2_1$. Assume three typical types of asteroids (as listed in Table~\ref{tab:1}), we calculate these values and plot the results in Fig.~\ref{fig:0}. Note the values (ordinate) is from $10^{-6}$ to $10^4$ covering a range of 10 orders of magnitude, implying that the values of $2, 2c_1, c_1^2$ differ greatly from each other under most circumstances. Therefore, generally there must be one dominant term among $2, 2c_1$ and $c^2_1$.

   \begin{table}
    \centering
    \caption{Thermal parameters for three typical types of asteroid, taken from \citet{far98}.}
    \label{tab:1}
    \begin{tabular}{l|cccc}
        \hline
         & $\rho$ (kg/m$^3$) & $K$ (W/m/K) & $C$ (J/kg/K) \\
        \hline
        Regolith-covered & 1500 & 0.0015 & 680 \\
        Basalt & 3500 & 2.65 & 680 \\
        Iron-rich & 8000 & 40 & 500 \\
        \hline
    \end{tabular}
   \end{table}

For $c_1\ll 1$, $c_1^2 \ll 2c_1 \ll 2$, the variations $\Delta a_s, \Delta a_d$ in equation~\eqref{eq:deltarg} become:
   \begin{eqnarray} \label{eq:deltar2}
    \begin{aligned}
     & \Delta a_s \approx -\frac{(\alpha
      L)^\frac{1}{4}(KC)^\frac{1}{2}a_0^{\frac{1}{4}}\sin^2\gamma}{6\sqrt{2}(\pi\mu\epsilon\sigma)^\frac{1}{4}\rho^\frac{1}{2}cR}t,\\
     & \Delta a_d \approx \frac{(\alpha
      L)^\frac{1}{4}(KC)^\frac{1}{2}a_0\omega_{\rm rot}^\frac{1}{2} \cos\gamma}{3\sqrt{2}(\pi\epsilon\sigma)^\frac{1}{4}(\mu\rho)^\frac{1}{2}cR}t.
    \end{aligned}
   \end{eqnarray}
If $c_1^2$ is dominant among $2, 2c_1, c^2_1$, similar calculations provide:
   \begin{eqnarray} \label{eq:deltar3}
    \begin{aligned}
     & \Delta a_s \approx -\frac{(\alpha L)^\frac{7}{4}(\epsilon\sigma)^\frac{1}{4}\sin^2\gamma}{24\sqrt{2}\pi^\frac{7}{4}\mu^\frac{3}{4}c\rho^\frac{3}{2}
      (KC)^\frac{1}{2}Ra_0^{\frac{5}{4}}}t,\\
     & \Delta a_d \approx \frac{(\alpha L)^\frac{7}{4}(\epsilon\sigma)^\frac{1}{4}\cos\gamma}{12\sqrt{2}\pi^\frac{7}{4}c\rho^\frac{3}{2} (\mu\omega_{\rm rot}KC)^\frac{1}{2}Ra_0^2}t.
    \end{aligned}
   \end{eqnarray}

It should be noted that the equations~\eqref{eq:deltar2} and \eqref{eq:deltar3} present the explicit dependence of $\Delta a$ on the thermal parameters at the cost of generality and accuracy. Therefore, we prefer to use equation~\eqref{eq:deltarg} to compute $\Delta a$ in practice. 

The estimations of the displacements of semimajor axis due to the Yarkovsky effect given in equations~\eqref{eq:deltar1}, \eqref{eq:deltar2} and \eqref{eq:deltar3} are all power functions of the parameters $K, C, \rho, R$ and $a_0$. These analytical results will be verified by numerical simulations in next section. It is worth noting that in between $R'\ll 1$ and $R'\gg 1$ a reasonable estimation for arbitrary $R'$ can be obtained by interpolating the two extreme situations.

\section{Numerical results}
Due to the seasonal and diurnal Yarkovsky effects, the orbital elements of an asteroid will evolve with time. The semimajor axis displacements $\Delta a_s, \Delta a_d$ depend on parameters such as the asteroid size $R$, density $\rho$, thermal conductivity $K$, specific heat capacity $C$, and initial semimajor axis $a_0$ (distance to the Sun). Particularly, these dependences are given explicitly in the analytical formulae that we obtained above.
In this section, we will first verify our analytical solutions by comparing them with the results derived by integrating perturbation equation and equation of motion. Then, the relation between the semimajor axis displacements $\Delta a_s, \Delta a_d$ and parameters $K, C, \rho, a_0$ are obtained and discussed.

\subsection{Calculation of $\Delta a_s$ and $\Delta a_d$}
The perturbation equations \eqref{eq:aeq} in Section~\ref{sec:pert} are often used to calculate $\Delta a_s$ and $\Delta a_d$. %Since the Yarkovsky effect sensitively depends on its radius $R$, here we first calculate this dependence. 
Assume three types of asteroids as summarized in Table~\ref{tab:1} and 1,000 different values of $R$ between $0.1$ to $10^4$\,m, we integrate equation~\eqref{eq:aeq} up to $10^7$\,yr to obtain the displacements in semimajor axis. The initial location $a_0$, period of rotation $P$, spin obliquity $\gamma$, absorption coefficient $\alpha$ and emissivity $\epsilon$ are set to be $a_0=2.5$\,AU, $\gamma=30^\circ$, $P=5$\,hours, and $\alpha=\epsilon=1$. 

   \begin{figure*}[htbp]
   \centering
   \includegraphics[width=\linewidth]{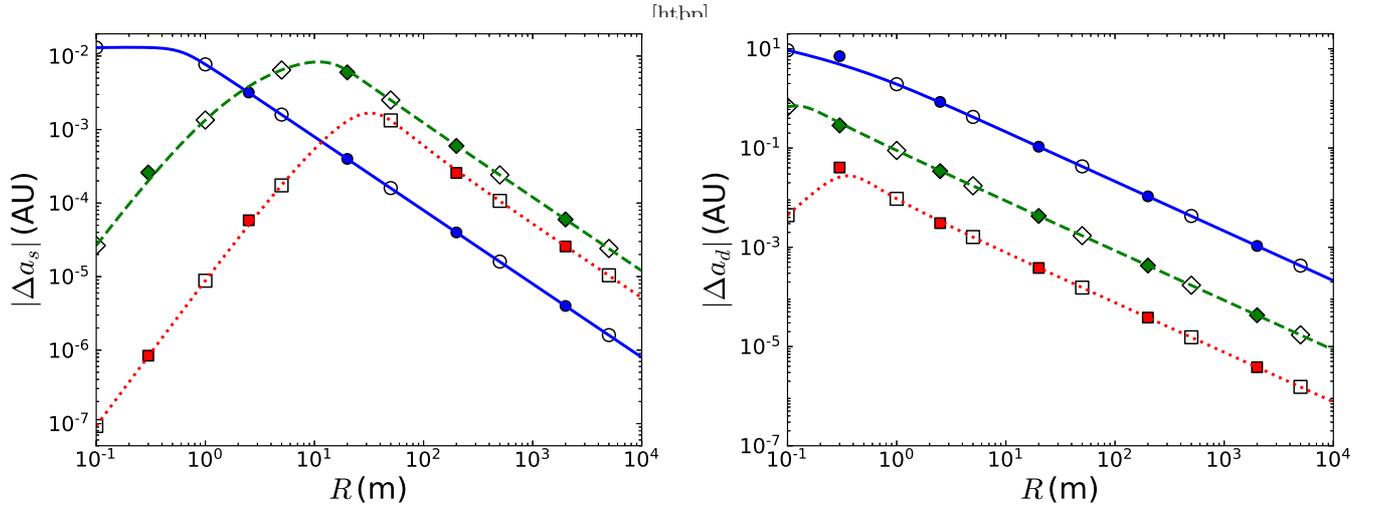}
    \caption{Displacements of semimajor axis $\Delta a_s$ (left panel) and
    $\Delta a_d$ (right panel) in $10^7$\,yrs versus the size of asteroid $R$. The lines are from the integration of perturbation equation~\eqref{eq:aeq}. Discrete points obtained by directly integrating the equation of motion (open points) and by analytical solutions (solid points) are plotted for comparison. The colours blue, green and red indicate types of asteroid respectively as regolith-covered, basalt and iron-rich (see Table~\ref{tab:1}). Note that the types of asteroid are also distinguished by styles of lines and points. }
    \label{fig:1}
   \end{figure*}

Without integrating the perturbation equation, which is relatively expensive in computation, we can also easily obtain the $\Delta a$ by directly computing the analytical formulae in equations~\eqref{eq:deltar1}, \eqref{eq:deltar2} and \eqref{eq:deltar3} derived previously in this paper. And certainly, the most reliable, but as well the most time-consuming way of finding $\Delta a$ is to integrate the equation of motion that includes the Yarkovsky force given in equation~\eqref{eq:force}. To do this, we simply consider a two-body problem consisting of the Sun and an asteroid that suffers the Yarkovsky effect. %The equation of motion of the asteroid is given in a Cartesian coordinate system and it is easy to transfer the force components in Eq.\,(\ref{eq:force2}) to the same coordinate system through a series of frame rotations. 
The asteroid's motion is numerically simulated using the integrator package {\it Mercury6} \citep{cham99} with a modification to include the Yarkovsky force, and its semimajor axis drift is regarded as the accurate result to verify the former two methods. The results are summarized in Fig.~\ref{fig:1}.

The lines in Fig.\,\ref{fig:1} obtained from the perturbation equation pass nearly exactly the open points that are the results of equation of motion, implying that equation~\eqref{eq:aeq} from the perturbation theory is perfectly reliable. The solid points calculated directly from the analytical estimations agree quite well with the lines (perturbation theory), thus also agree with the reality (equation of motion), verifying the reliability of our analytical formulae in equations~\eqref{eq:deltar1}, \eqref{eq:deltar2} and \eqref{eq:deltar3}.

The Yarkovsky effect may be significant. As we can see from Fig.~\ref{fig:1}, the maximal semimajor axis migration $\Delta a_s$ can be $\sim$0.01\,AU in $10^7$\,yr, and $\Delta a_d$ can reach $\sim$10\,AU. Except for the regolith-covered type asteroids that have a very low thermal conductivity $K$, the asteroids of tens of meters drift the most due to the seasonal effect, and those of decimetres do due to the diurnal effect. The $\Delta a$ is relatively small for both very small and very large asteroids. In the log-log scale plot of Fig.~\ref{fig:1}, $\log\Delta a$ changes linearly with respect to $\log R$ for both ends of small and large sizes. The slopes of the lines ($+2$ on the left side and $-1$ on the right side) for both seasonal and diurnal effects coincide very well with the analytical results, i.e. $\Delta a\propto R^2$ for $R'\ll 1$ in equation~\eqref{eq:deltar1} and $\Delta a\propto R^{-1}$ for $R'\gg 1$ in equations~\eqref{eq:deltar2} and \eqref{eq:deltar3}. Apparently, the normalized sizes $R'_s$ and $R'_d$ of the asteroid depend on the scales (penetration depth $l_s$ and $l_d$), thus the conditions $R'\ll 1$ and $R'\gg 1$ differ significantly from each other for the seasonal and diurnal effects.

\subsection{Dependence on thermal parameters}

\begin{figure*}%[htbp]
   \centering
   \includegraphics[width=\linewidth]{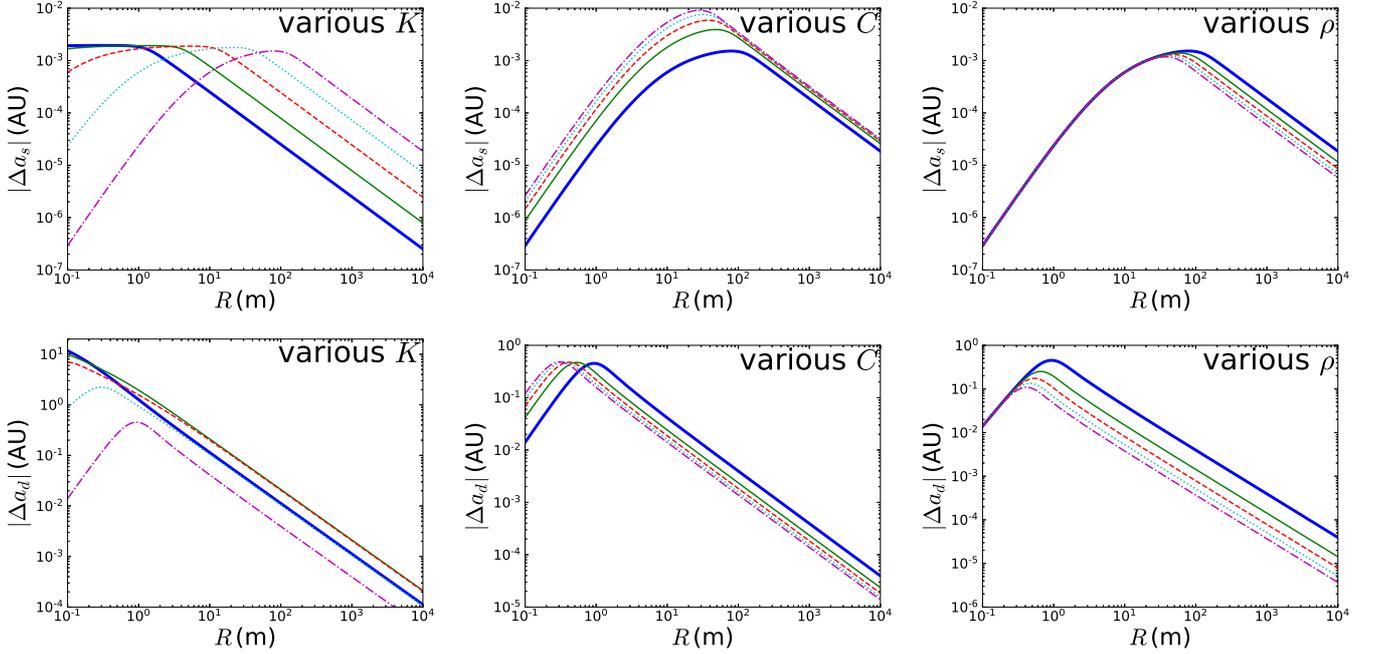}
   \caption{The Yarkovsky effect of asteroids with different thermal parameters in $10^7$\,yr. The upper and lower rows are for seasonal and diurnal effects. In each panel, two parameters are fixed while the third one (labelled in the panel) is specified to five values. The curves in blue, green, red, light blue
   and purple are assigned to represent these five cases in sequence as follows. Left: $C=100$, $\rho=1500$ and $K=0.001, 0.01, 0.1, 1, 10$. Middle: $K=10, \rho=1500$ and $C=100, 300, 500, 700, 900$. Right: $K=10, C=100$ and $\rho=1500, 3000, 4500, 6000, 7500$.
   The units of $K, C, \rho$ are the same as Table~\ref{tab:1}.}
   \label{fig:2}
 \end{figure*}

The Yarkovsky effect is strongly influenced by the thermal properties of the asteroid, as indicated by the different ranges and profiles of three curves for different types of asteroids in Fig.~\ref{fig:1}. These dependences of Yarkovsky effect on the thermal parameters have been revealed explicitly by the analytical formulae in equations~\eqref{eq:deltar1}, \eqref{eq:deltar2} and \eqref{eq:deltar3}. To illustrate these dependences over a wide range of parameter values, we numerically integrate the perturbation equation, i.e. equation~\eqref{eq:aeq}, and compute the displacement $\Delta a$ of asteroids of varying thermal parameters and different sizes initially located at $a_0=2.5$\,AU. The results after $10^7$\,yr's evolution, summarized in Fig.~\ref{fig:2}, show clearly how the thermal conductivity $K$, specific heat capacity $C$ and the bulk density $\rho$ influence the Yarkovsky effect. And these dependences are consistent with the explicit relation given in equations~\eqref{eq:deltar1}, \eqref{eq:deltar2} and \eqref{eq:deltar3}. 

Checking the dependence of the displacements $\Delta a_s, \Delta a_d$ on $K, C$ and $\rho$ given by the analytical estimations in Section~\ref{sec:est}, we find
 \begin{equation} \label{eq:estsmall}
\Delta a_s, \Delta a_d \propto K^{-2}C^1\rho^0
 \end{equation}
in equation~\eqref{eq:deltar1} for small asteroids ($R'\ll 1$). And for large asteroids ($R'\gg 1$), the dependence is
\begin{equation} \label{eq:estbig1}
\Delta a_s, \Delta a_d \propto K^{1/2}C^{1/2}\rho^{-1/2}, \hspace{0.5 cm} \text{if } c_1^2\ll 2
\end{equation}
as in equation~\eqref{eq:deltar2}, or 
\begin{equation} \label{eq:estbig2}
\Delta a_s, \Delta a_d \propto K^{-1/2}C^{-1/2}\rho^{-3/2}, \hspace{0.5 cm} \text{if } c_1^2\gg 2
\end{equation}
as in equation~\eqref{eq:deltar3}. When $a_0=2.5$\,AU, the condition $(c_1)_s^2\gg 2$ for seasonal effect can be met only if the ``thermal inertia'' $(\rho KC)$ is very large (see equation~\eqref{eq:c12def} and Fig.~\ref{fig:0}). Thus, for parameters we adopted here to plot Fig.~\ref{fig:2}, the dependence as in equations~\eqref{eq:estsmall} and \eqref{eq:estbig1} should be applied for seasonal effect $\Delta a_s$. However, for diurnal effect, since $(c_1)_d^2$ is hardly much less than 2, equation~\eqref{eq:estbig2} should be employed for $\Delta a_d$ in most cases.

Bearing this in mind, we know that as $K$ increases, the left end of a curve (small $R$, for either $\Delta a_s$ or $\Delta a_d$) in Fig.~\ref{fig:1} will go down as indicated by equation~\eqref{eq:estsmall} and the right end (large $R$) of $\Delta a_s$ will rise up as indicated by equation~\eqref{eq:estbig1}. Meanwhile, the right end of $\Delta a_d$ increases a little first and then drops down following equation~\eqref{eq:estbig2}. The gradients of the curve in the far-left end and far-right end in Fig.~\ref{fig:1} remain constants (i.e. $+2$ and $-1$ respectively), which is equivalent to a shift toward the right of the curve such that the maximal displacement ($\Delta a_{\max}$) will attain at a larger size $R$. This is exactly what one can see in the left panels of Fig.~\ref{fig:2}.

As $C$ increases, the left end of a curve will go up, but as for the right end, the curves increase for $\Delta a_s$ and they drop down for $\Delta a_d$, which is consistent with what the middle panel of Fig.~\ref{fig:2} shows.

Finally, for small asteroids, the $\Delta a_s$ and $\Delta a_d$ displacements are independent of the density $\rho$ , while for large asteroids $\Delta a_s$ and $\Delta a_d$ drop down as $\rho$ increases. The right panel of Fig.~\ref{fig:2} behaves in an analogous way.

\subsection{Dependence on initial position}

\begin{figure*}%[htbp]
    \centering
    \includegraphics{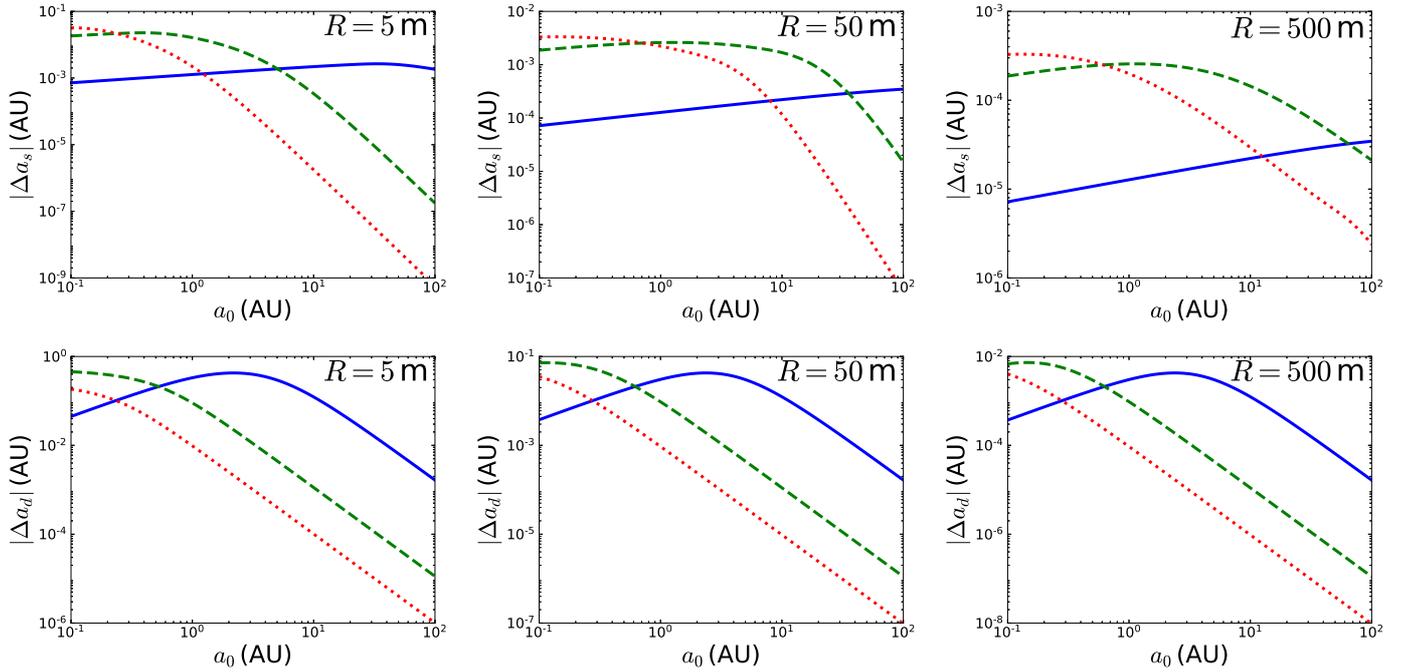}
    \caption{Displacements of semimajor axis of seasonal effect $\Delta a_s$ (upper panels) and diurnal effect $\Delta a_d$ (lower panels) in $10^7$\,yr versus the initial semimajor axis $a_0$. Three different types of objects, of regolith-covered (blue solid), basalt (green dashed) and iron-rich (red dotted), are considered. The left, middle and right panels are for objects of size $R=5, 50$ and $500$\,m, respectively. }
    \label{fig:3}
   \end{figure*}

According to its distance to the Sun, an asteroid receives different radiation flux, subsequently the Yarkovsky effect changes with the semimajor axis $a_0$. Such a dependence has been mentioned briefly in Section~\ref{sec:pert} and it can also be found in equations~\eqref{eq:deltar1}, \eqref{eq:deltar2} and \eqref{eq:deltar3}. However, theoretically the estimations in these formulae are valid only when either $R'\ll 1$ or $R'\gg 1$ and when $\Delta a_s, \Delta a_d$ are small. These conditions are not necessarily met in practice. Therefore, it is still beneficial to calculate numerically $\Delta a_s$ and $\Delta a_d$ at different $a_0$ by integrating the perturbation equation~\eqref{eq:aeq} and then verify the applicability of the analytical formulae. 
For three typical types of asteroids as listed in Table~\ref{tab:1} and four typical sizes of asteroids $R=5, 50, 500$\,m and $5$\,km, we calculate numerically $\Delta a_s$ and $\Delta a_d$ at different $a_0$ ranging from $0.1$ to $100$\,AU. The results for $R=5, 50, 500$\,m are plotted in Fig.~\ref{fig:3}. The plots for $R=5$\,km are omitted because the profiles of curves for this case are just the same as the ones for $R=500$\,m, only except the displacements of both $\Delta a_s$ and $\Delta a_d$ for the former are one order of magnitude lower than the latter.

From Fig.~\ref{fig:3}, we see that the Yarkovsky effect changes dramatically as the distance from the Sun increases, and the profiles of curves for different types of asteroids may be quite different. Because the approximation conditions for the analytical estimations in equations~\eqref{eq:deltar1}, \eqref{eq:deltar2} and \eqref{eq:deltar3} are defined by the scaled size $R'$ that depends on the thermal penetration depth, again, asteroids of the same physical size but with different compositions may meet either the condition $R'\ll 1$ or $R'\gg 1$ in most cases.

The upper three panels of Fig.~\ref{fig:3} show the seasonal effect. In the left and middle panels, the right end of the curves for the iron-rich asteroids (high $K$ and $\rho$, represented by red dotted lines) can be fitted well by straight lines. This linear relation in the log-log scale means a power law $\Delta a_s \propto a_0^k$. The slope for $\Delta a_s$ is $k=-7/2$, perfectly consistent with the exponent in equation~\eqref{eq:deltar1}, implying that the condition $R'_s\ll 1$ is fulfilled for such $5$\,m and $50$\,m in size, iron-rich asteroids.
On the contrary, in these two top panels, the blue solid line, representing the regolith-covered objects with very small thermal conductivity $K$ thus very small thermal penetration depth, does not follow the same law as the iron-rich object does. In fact, for such regolith-covered object with a ``heat-insulated surface'', even a size of 5\,m is not ``small''.

For objects of $500$\,m in size (or bigger), they are large ($R'_s\gg 1$). Note in Fig.~\ref{fig:0} that $(c_1)_s^2\gg 2(c_1)_s \gg 2$ for iron-rich objects at distance farther than several AUs, thus the estimations in equation~\eqref{eq:deltar3} should be applied. In the right panels of Fig.~\ref{fig:3} the slope of the red dotted line representing the iron-rich objects at the right end is around $-1$ for $\Delta a_s$, matching quite well with the exponents ($-5/4$) in equation~\eqref{eq:deltar3}. 
For regolith-covered objects, $2\gg 2(c_1)_s \gg (c_1)_s^2$ for all $a_0$ (see Fig.~\ref{fig:0}), thus equation~\eqref{eq:deltar2} should be used to calculate the $\Delta a_s$, and the slopes of $1/4$ (for $\Delta a_s$) of the blue solid lines in the right  panels of Fig.~\ref{fig:3} are perfectly consistent with the theoretical expectations in equation~\eqref{eq:deltar2}. Note that the positive slope $1/4$ means that for such kind of objects the seasonal effect drives the outer bodies to migrate inward by bigger distance than the inner ones.

The case of basalt asteroids is in between cases of the regolith-covered asteroids and the iron-rich asteroids.

For the diurnal effect shown in the lower three panels of Fig.~\ref{fig:3}, the right end of all the curves can also be well fitted by straight lines. The slope $k=-2$ for $\Delta a_d$ is perfectly consistent with the exponent in equations~\eqref{eq:deltar1} and \eqref{eq:deltar3}.
Note that in the lower left panel of Fig.~\ref{fig:3}, the curve for the iron-rich asteroids is consistent with equation~\eqref{eq:deltar1}, while the other curves in the lower three panels are consistent with equation~\eqref{eq:deltar3}, although they have the same slope value of $-2$.

Perhaps the most interesting phenomenon presented in Fig.~\ref{fig:3} is that the displacement $\Delta a$ does not decrease monotonically as the distance to the Sun increases. In fact, besides the solar radiation flux that is obviously anti-correlated with the semimajor axis, the Yarkovksy effect depends also on temperature distribution on the object's surface that is reflected by the thermal parameter $\Theta$ as shown in equation~\eqref{eq:relax}, resulting finally in a  complicated dependence on semimajor axis. As we have mentioned in Section~\ref{sec:pert}, for diurnal effect the dependence may change from $\dif a/\dif t \propto a$ to $\dif a/\dif t \propto a^{-2}$ as $\Theta_d$ increases. The maximal $\Delta a_d$ will be met at some $\Theta_d$.

For large asteroids satisfying $R'\gg 1$, substitute equation~\eqref{eq:cofrb} into \eqref{eq:aeq}, we may find that $\dif a / \dif t$ is a function of $c_1$ and $c_2$, which are given in equation~\eqref{eq:c12def}. For simplicity, denote $(c_1)_d=p_1 a^{3/2}$ and $(c_2)_d=p_2 a^{-2}$ after equation~\eqref{eq:c12def}, where $p_1, p_2$ are constants, and easy calculation leads to
\begin{equation}
\left(\frac{\dif a}{\dif t}\right)_d \propto \frac{p_1p_2a}{2+2p_1a^{3/2}+p_1^2a^3},
\end{equation}
and
\begin{equation}
\frac{\dif}{\dif a}\left(\frac{\dif a}{\dif t}\right)_d \propto \frac{2-p_1a^{3/2}-2p_1^2a^3}{(2+2p_1a^{3/2}+p_1^2a^3)^2}.
\end{equation}
Thus, the maximum of $\dif a / \dif t$ appears when $2-p_1a^{3/2}-2p_1^2a^3=0$, i.e. $p_1a^{3/2}\approx 0.78$. The substitution of the parameters adopted in this paper yields the maximal $\Delta a_d$ at $a=2.4$\,AU, 0.15\,AU and 0.051\,AU for the regolith-covered, basalt and iron-rich asteroids, respectively. We note that these values are consistent with the results in Fig.~\ref{fig:3}, and for the main belt asteroids $\dif a / \dif t\propto a^{-2}$ is still a good estimation in most cases. Similar calculations can be conducted for $R'\ll 1$ and for seasonal effect.

\section{Diurnal and seasonal effects}

%\subsection{Combination of the seasonal and diurnal effects}
In practice, the Yarkovsky effect is a combination of the seasonal and diurnal effects (and the mixed term that is negligible, see \cite{vok99}). For a typical small body whose spinning rate is much faster than the mean motion, if the obliquity $\gamma$ is not around $90^\circ$, the diurnal effect is much larger than the seasonal effect, so the latter generally can be just neglected. Consequently, the combined semimajor shift $\Delta a_{\rm total} \approx \Delta a_d$. But the seasonal effect may become comparable to the diurnal effect as the size $R$ increases, and in this instance, both of them shall be taken into account simultaneously, i.e. $\Delta a_{\rm total} \approx \Delta a_s+\Delta a_d$.

The semimajor axis displacements $\Delta a_s$ and $\Delta a_d$ are proportional to $-\sin^2\gamma$ and $\cos\gamma$ respectively, therefore the seasonal effect always pushes the body inward no matter what the spin obliquity $\gamma$ is, while the direction of diurnal effect depends on $\gamma$. When the spin is prograde (retrograde), the diurnal effect pushes the celestial body outward (inward). %In addition, the strength of these two effects varies with different values of $R$, thus an asteroid suffering the Yarkovsky effect may migrate inward or outward. 

For a retrograde spinning body, since the seasonal and diurnal effects are in the same direction, it will migrate inward (toward the Sun) due to the Yarkovsky effect, obviously. 

A prograde spinning body has a more complex story. Depending on the obliquity and spin rate, the direction of migration may change. Since $\Delta a_s,\Delta a_d$ have been analytically given in equations~\eqref{eq:deltar1}, \eqref{eq:deltar2} and \eqref{eq:deltar3}, we may use these formulae to find when the migration changes direction.

For small asteroids satisfying $R'\ll 1$, the total semimajor displacement can be estimated through equation~\eqref{eq:deltar1},
\begin{equation}
    \Delta a_s + \Delta a_d = \frac{(\alpha L)^\frac{7}{4}(\epsilon\sigma)^\frac{1}{4}CR^2}{120\pi^\frac{7}{4}K^2ca_0^\frac{7}{2}} \left(\frac{2\omega_{\rm rot}}{\omega_{\rm rev}}\cos\gamma-\sin^2\gamma\right)t.
\end{equation}
Denote 
\begin{equation}
    \beta=\frac{\omega_{\rm rot}}{\omega_{\rm rev}}, 
\end{equation}
and we know that the asteroid will migrate outward if
\begin{equation}
\left(2\beta\cos\gamma-\sin^2\gamma\right) > 0,
\end{equation}
or alternatively,
\begin{equation} \label{eq:criterion}
    \cos\gamma > \sqrt{1+\beta^2}-\beta.
\end{equation}
Generally the spin is much faster than the revolution, $\omega_{\rm rot} \gg \omega_{\rm rev}$, thus $\beta \gg 1$ and the inequality turns to 
\begin{equation} \label{eq:crit1}
    \cos\gamma > \sqrt{1+\beta^2}-\beta= \beta\sqrt{1+1/\beta^2}-\beta\approx \frac{1}{2\beta}. 
\end{equation}
This condition can be fulfilled for almost all prograde obliquity, only except for the case when $\gamma$ is very close to $90^\circ$. 

If $R' \gg 1$, however, the $\Delta a_s$ and $\Delta a_d$ may be calculated using the analytical estimations given in equation~\eqref{eq:deltar2} or equation~\eqref{eq:deltar3}, depending on the value of parameters $(c_1)_s$ and $(c_1)_d$. Since $(c_1)_d=\sqrt{\beta} (c_1)_s$ and generally $\beta \gg 1$, we may have 
\begin{enumerate}
	\item $(c_1)_s \ll (c_1)_d \ll 1$, or
	\item $1 \ll (c_1)_s \ll (c_1)_d$, or
	\item $(c_1)_s \ll 1 \ll (c_1)_d$. 
\end{enumerate}
For case (i), both $\Delta a_s$ and $\Delta a_d$ are given by equation~\eqref{eq:deltar2} and simple algebraic calculation leads to the condition for outward migration
\begin{equation} \label{eq:crit2}
    \cos\gamma > \sqrt{1+\beta}-\sqrt{\beta}\approx \frac{1}{2\sqrt{\beta}}.
\end{equation}

For Case (ii), equation~\eqref{eq:deltar3} is applied for both $\Delta a_s,\Delta a_d$ and the outward migration happens when 
\begin{equation} \label{eq:crit3}
    \cos\gamma > \sqrt{1+\frac{1}{\beta}} - \sqrt{\frac{1}{\beta}}. 
\end{equation}

For Case (iii), take $\Delta a_s$ from equation~\eqref{eq:deltar2} and $\Delta a_d$ from equation~\eqref{eq:deltar3}, and $\Delta a_s + \Delta a_d > 0$ requires 
\begin{equation} \label{eq:crit4}
    \cos^2\gamma + \frac{1}{\mu\rho KC}\sqrt{\frac{\epsilon\sigma(\alpha L)^3 \omega_{\rm rev}^3}{4\pi^3\omega_{\rm rot}}}\cos\gamma -1 >0.
\end{equation}

All the criteria in equations~\eqref{eq:crit1}--\eqref{eq:crit4} tell the same fact that the migration direction is determined by the obliquity and the spinning rate (and the thermal parameters as well). As soon as the thermal parameters are given, we can choose the appropriate criterion and determine the migration direction. For instance, a regolith-covered  asteroid of size $R=50$\,m (thus $R'\gg 1$) and rotation period $P=5$\,hours in near circular orbit at 2.5\,AU (thus $\beta= 6,930\gg 1$) has $(c_1)_s=0.01$ and $(c_1)_d=0.83$, therefore the criterion in equation~\eqref{eq:crit2} should be applied, and calculation shows that it migrates outwards if $\gamma < 89^\circ$. For iron-rich and basalt asteroid, similar calculations show that the criteria in equations~\eqref{eq:crit3} and \eqref{eq:crit4} should be employed and the critical obliquities for these types of asteroid are $9^\circ$ and $19^\circ$, respectively.

\begin{figure*}%[htbp]
	\centering
	\includegraphics{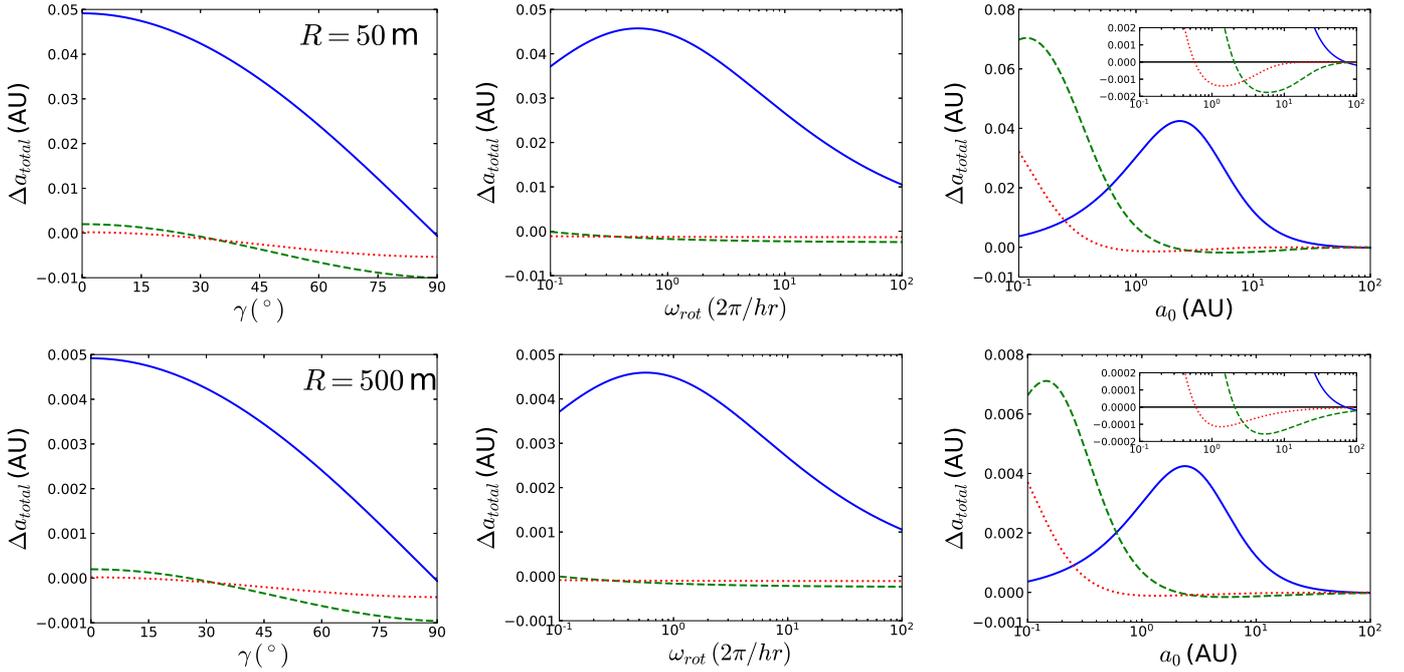}
	\caption{Displacements of semimajor axis $\Delta a_{\rm total}$ in $10^7$\,yr versus the obliquity $\gamma$ (left panel, for $a_0=2.5$\,AU and the rotation period $P=5$\,hours), the rotation frequency $\omega_{\rm rot}$ (middle panel, for $a_0=2.5$\,AU,  $\gamma=30^\circ$) and the initial semimajor axis $a_0$ (right panel, for $P=5$\,hours,  $\gamma=30^\circ$). Sizes of $R=50$\,m (upper panels) and $R=500$\,m (lower panels) are assumed, and three different types of objects, of regolith-covered (blue solid), basalt (green dashed) and iron-rich (red dotted), are considered. The embedded window in the right panel shows the zero points (intersections of curves and the black line of $\Delta a_{\rm total}=0$). }
	\label{fig:6}
\end{figure*}

For the sake of obtaining an overall knowledge about the migration direction and extent, we numerically integrate the perturbation equation and find the total semimajor axis displacement $\Delta a_{\rm total}$ under both the seasonal and diurnal Yarkovsky effects as the function of obliquity $\gamma$, spinning rate $\omega_{\rm rot}$ and initial semimajor axis $a_0$. Two sizes of $R=50$\,m and $R=500$\,m are assumed and the equations are integrated to $10^7$\,yr. The results are plotted in Fig.~\ref{fig:6}. As shown in the left panel of Fig.~\ref{fig:6}, the Yarkovsky effect pushes the regolith-covered asteroid outward for all $\gamma \lesssim 89^\circ$. As for the iron-rich asteroid, it migrates very little outward when $\gamma \lesssim 12^\circ$ and goes inward otherwise. The basalt asteroid however, migrates a little further than the iron-rich one, both outward and inward, with the turning point at $\gamma\approx 25^\circ$. These critical points of $\gamma$ are basically consistent with the values obtained through the analytical estimation given above. 

The dependence of $\Delta a_{\rm total}$ on the rotation rate $\omega_{\rm rot}$ (thus $\beta$) is shown in the middle panel of Fig.~\ref{fig:6}. For the regolith-covered asteroid, $\Delta a_{\rm total}$ first rises then declines with increasing $\omega_{\rm rot}$, because the appropriate analytical estimation for $\Delta a_d$ changes from equation~\eqref{eq:deltar2} to equation~\eqref{eq:deltar3} as the parameter $(c_1)_d$ increases with $\omega_{\rm rot}$, and consequently the relationship $\Delta a_d \propto \omega_{\rm rot}^{1/2}$ turns to $\Delta a_d \propto \omega_{\rm rot}^{-1/2}$. For the iron-rich asteroid, the diurnal effect $\Delta a_d \propto \omega_{\rm rot}^{1/2}$ is always applied, but the seasonal effect may dominate sometimes, so we cannot see very obvious variation of the total Yarkovsky effect with increasing $\omega_{\rm rot}$. Finally, the curve for the basalt asteroid is in between the above two cases. 

Although the dependence of $\Delta a_{\rm total}$ on the initial semimajor axis $a_0$ can be obtained by adding the corresponding $\Delta a_s$ and $\Delta a_d$ in Fig.~\ref{fig:3}, we still plot this relation in the right panel of Fig.~\ref{fig:6} to emphasize that there are ``zero points'' in these curves. All the three curves drop from positive to negative and cross the zero point ($\Delta a_{\rm total}=0$) at some $a_0$. Obviously, asteroids with $\Delta a_{\rm total}>0$ on the left side of the zero point will migrate outward, and vice versa. Therefore, asteroids will accumulate around such zero points. These points locate at $0.59, 2.0$ and $72$\,AU respectively for three curves. 

There exists another kind of zero points, through which the curve increases from negative to positive. Contrary to the previous zero points, asteroids will diverge from these positions, leaving a gap there. Theoretically, these two kinds of zero points may form structures among the disk of asteroids or planetesimals by piling up them in some locations and depleting them somewhere else. The location of these zero points surely depend on the asteroids' physical, thermal and dynamical parameters, thus they provide a sort of ``selection'' mechanism in forming the space distribution for asteroids with specific parameters. 

We note that all the locations of the turning points or zero points on the $\gamma, \omega_{\rm rot}$ or $a_0$ axis for the object of $R=50$\,m are the same as the ones for $R=500$\,m. This is not strange since all the criteria in equations~\eqref{eq:crit1}--\eqref{eq:crit4} are independent of the size $R$. In fact, as long as $R'\gg 1$ is satisfied, both $\Delta a_s$ and $\Delta a_d$ are proportional to $R^{-1}$ as shown in equations~\eqref{eq:deltar2} and \eqref{eq:deltar3}, thus the zero points are the same for these two sizes but the migration distance for $R=500$\,m is one order of magnitude smaller than that of $R=50$\,m, just as shown in Fig.~\ref{fig:6}. 

The time needed for an asteroid drifting from its initial position to the zero point can be estimated as follows. Simply assume that the semimajor axis drift speed $\dif a/\dif t$ of an asteroid initially at $a_0$ depends linearly on its distance to the zero point $a_z$ 
\begin{equation}
    \frac{\dif a}{\dif t} = \left(\frac{\dif a}{\dif t}\right)_{a=a_0}\times \frac{a-a_z}{a_0-a_z},
\end{equation}
which indicates
\begin{equation}
    \left|\frac{a-a_z}{a_0-a_z}\right| = \mathrm{e}^{-t/t_z},
\end{equation}
where $t_z$ is the time scale to reach the zero point, and 
\begin{equation}
    t_z = \frac{a_z-a_0}{\left(\dif a/\dif t\right)_{a=a_0}}.
\end{equation}
Generally, this time scale is so long ($\sim$10\,Gyr) that the orientation of spin axis and rotation rate of an asteroid may be greatly changed by the YORP effect \citep{cap04}. As a result, such balance position can hardly be reached by an asteroid in reality unless it is originally very close to the balance position and it is coincidentally in the equilibrium phase of the YORP cycle, which is very unlikely to be true. Even so, the mutual collisions among the disk will destroy the balance.

\section{Distribution of Eos family}

With the help of the analytical solutions derived in Section~\ref{sec:est} we have analysed the dependence of the semimajor axis drift due to Yarkovsky effect on the physical, thermal and dynamical parameters of asteroids. These dependences are explicitly presented in the analytical solutions, making the calculation very easy if all the parameters are given. On the other hand, some of these parameters, for example the thermal parameters and the rotation state of asteroids, cannot be precisely determined through the limited observations. In this case, the explicit solutions can be easily applied to test and adjust these parameters so that the theoretical expectations can match the observations. As an example of such application of the analytical estimations, we investigate below the space distribution of the members of the Eos family. 

\subsection{Determination of parameters}
\begin{figure}%[htbp]
\centerline{
\includegraphics[width=9.0cm,height=7.50cm,angle=0]{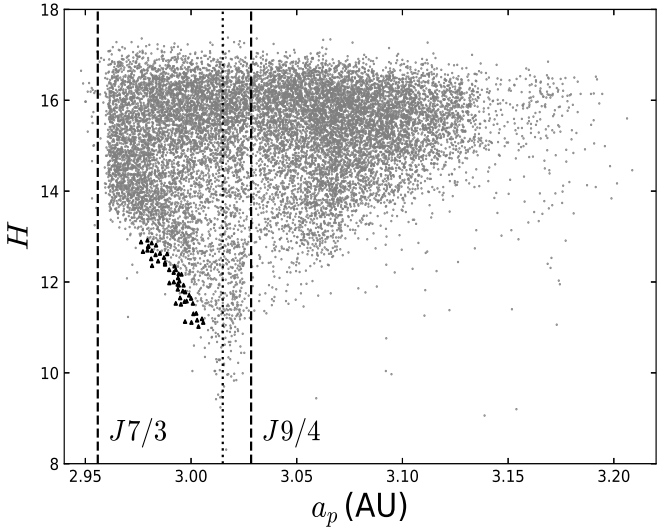}}
\caption{Distribution of the Eos family members in the $H-a_p$ plane, where $H$ is magnitude and $a_p$ is proper semimajor axis. The centre of the family is indicated by the dotted line, while the positions of the 7/3 and 9/4 MMR with Jupiter are denoted by dashed lines. Black triangles are the members used to calibrate the thermal conductivity $K$ (see text).} 
\label{fig:8}
\end{figure}

The Eos family is a prominent asteroid family with the number of recognized members around 15,000. We download from {\it AstDyS} website\footnote{http://newton.spacedys.com/astdys/} the data (orbital elements and absolute magnitude) of 14,785 asteroids that are labelled as Eos family members \citep{mil14} and plot them in Fig.~\ref{fig:8}.
For comparison, the same number of test asteroids with the sizes estimated by the absolute magnitude $H$ and a geometric albedo $p_v=0.13$ \citep{vok06} are generated. 

Since all the family members are fragments of the same parent body initially at $3.015$\,AU and smashed $1.3$\,Gyr ago \citep{hanus18}, it is reasonable to assume that they have the same bulk density and thermal parameters. In literatures, both \citet{vok06} and \citet{broz13} suggest the bulk density $\rho=2500\,$kg/m$^3$ and the specific heat capacity $C=680$\,J/kg/K, but they give a little different values for thermal conductivity, $K=0.005$\,W/m/K and $K=0.001$\,W/m/K, respectively. For the Bond albedo, \citet{broz13} suggests a value $A=0.1$, thus the absorption coefficient $\alpha=1-A=0.9$.

The rotation rate $\omega_{\rm rot}$ is independent of the obliquity $\gamma$, while $\omega_{\rm rot}$ could depend on the asteroid size $R$, which in turn can be estimated through its photometric magnitude (brightness) after assuming a geometric albedo. During the fragmentation event that generated this asteroid family, if the rotational kinetic energy is uniformly distributed to the mass, i.e. $I\omega_{\rm rot}^2\propto m$, where $I\propto mR^2$ is the moment of inertia, then qualitatively, $\omega_{\rm rot} \propto R^{-1}$. Simply assume $\omega_{\rm rot}=b/R$, in which $b$ is a pending constant, and we test several $b$ values. We finally set $b=0.502$\,m/s (where $R$ in metres), which makes the distribution of $\omega_{\rm rot}$ match the observation results in \citet{warner09} best. Of course, here we do not deny the possibility of other relations between $\omega_{\rm rot}$ and $R$. In fact, we also test some other relations later in this paper, but we note that $\omega_{\rm rot}=0.502/R$ gives the best result. 

We select randomly several asteroids located on the edge of the V-plot (marked by black triangles in Fig.~\ref{fig:8}), which should have migrated farthest, and use their migration distances to calibrate the conductivity $K$. Adopt the above mentioned $\rho=2500$\,kg/m$^3$ and $C=680$\,J/kg/K, and assume $\gamma=0^\circ$ or $180^\circ$ (this obliquity causes maximal migration). The spin rate is estimated using $\omega_{\rm rot}=0.502/R$ (the size $R$ is calculated from $H$) if $\omega_{\rm rot}$ is not available in the observation data. Set $K$ in the possible domain and then we calculate the migration distances of these objects from the original site ($a_0=3.015$\,AU). Comparing the results with their current positions, we get the best values of $K$ for each of these fringe objects that make the positions match best. We derive their average value of $K=0.008$\,W/m/K, and adopt it below in our calculations. Be aware of that there is a large uncertainty in $K$ \citep{hanus18b}, and we note that this value is basically consistent with the previous estimation \citep{vok06,broz13}. 

\subsection{Distributions of $\gamma$ and $\omega_{\rm rot}$}
To complete the calculation of the migration of each family member, we should have the full knowledge of the spin state, that is, the distributions of obliquity $\gamma$ and spin rate $\omega_{\rm rot}$. 

A brief review of current distributions of obliquity $\gamma$ and rotation frequency $\omega_{\rm rot}$ of Eos family members can be found in \citet{hanus18}. For $\gamma$, a model is established in \citet{cib16} to derive the spin orientation with photometric data sparse in time. Obtained from a theoretical model, this distribution of $\gamma$ perhaps does not exactly agree with the statistical results of real members. The distribution of $\omega_{\rm rot}$ is calculated based on observational data in \citet{warner09}. Although the total number of samples in this study is just 569, about 1/25 of the known Eos family members, it is still the most probable distribution we could have at present. 

Right after the event that produced the Eos family, most probably the initial distribution of spin orientation in space is isotropic. Equivalently, an isotropic orientation means that the value of $\cos\gamma$ distributes randomly in the range $[-1, 1]$. The subsequent collisions among the members may change the spin states, but the distribution will not be affected if the collisions happen randomly. The YORP effect however, may bring some modification to the distribution of $\cos\gamma$, because it always leads the obliquity $\gamma$ to $0^\circ$ or $180^\circ$ \citep{vokcap02}. Obviously, how much the distribution deviates from a random $\cos\gamma$ depends on the time scales of the collisions and of the YORP effect. In this paper, we will not discuss the complex evolution of obliquity of family members. Instead, we simply test some other distributions, among which a random obliquity distribution ($\gamma$ randomly locates inside $[0^\circ, 180^\circ]$) is of special interest. Compared to the former distribution (random $\cos\gamma$), in this distribution (random $\gamma$) relatively more family members have their obliquities close to $0^\circ$ or $180^\circ$, roughly reflecting the modification by YORP effect statistically. Lastly, we also test the distribution obtained from the limited observations \citep{cib16}. 

Similar arguments as above can be applied to the distribution of spin rates $\omega_{\rm rot}$ of family members. For simplicity, we assume that $\omega_{\rm rot}$ depends only on the size $R$ (but not $\gamma$) via $\omega_{\rm rot} \propto R^{-k}$. We have adopted $k=1$ for determining the $K$ value in last subsection, and we will test other values of $k$ later. And we also test the possibility that $\omega_{\rm rot}$ is independent of $R$. In this case, we just assign to each test member a constant $\omega_{\rm rot}$ or a random $\omega_{\rm rot}$ which follows the distribution given by the 569 observation samples \citep{warner09}. 

\subsection{7/3 and 9/4 mean motion resonances with Jupiter}
Two major mean motion resonances (MMR) with Jupiter, 7/3 MMR at $2.957$\,AU and 9/4 MMR at $3.03$\,AU, interrupt the region of the Eos family. The MMR may pump up the eccentricity and cause ejection of asteroids, contributing to the formation of their final space distribution. The 7/3 MMR is so strong that it will eliminate nearly all asteroids passing through \citep{tsi03}. As a result, it serves as a sharp boundary of the Eos family in practice. Therefore, in our calculations, we just remove those objects that reach this resonance through migration via Yarkovsky effect. 

The 9/4 MMR is relatively weak, and thus cannot eliminate asteroids as efficiently as the 7/3 MMR. In fact, the motion in/around this MMR is quite complicated \citep{mor95}. To estimate the efficiency of the 9/4 MMR in depleting asteroids, we compare the numbers of family members in both sides of the family centre, as follows. We note a similar method has been applied in \citet{vok06}. 

Suppose the original family members diffuse inward and outward from the centre symmetrically. Thus we expect to see the same amount of members in a given range of region in both left and right hand sides, except that this symmetry is broken by some mechanisms. The 7/3 MMR in the left hand side and the 9/4 MMR in the right hand side, are such asymmetric mechanisms. In between the 7/3 MMR and the family centre, the region from $2.965$\,AU to $3.015$\,AU of width $0.05$\,AU, in which no other major resonance exists, hosts 5036 Eos family members. On the right hand side, the region from $3.015$\,AU to $3.065$\,AU of the same width ($0.05$\,AU) hosts only 4476 members. The deficiency of 560 $(=5036-4476)$ members may be attributed mainly to the depletion of the 9/4 MMR, although some high-order MMRs and three-body resonances can be found in both of the above mentioned regions \citep{vok06} and some young subfamilies produced by secondary collisions may affect the number of objects in each interval \citep{tsi19}. We simply adopt $560/5036=11\%$ as the probability of depletion by the 9/4 MMR. 

Generally, the slower an asteroid migrates, the more possible it will be affected by the 9/4 MMR. Therefore, we transfer this probability of depletion to a migration speed at the resonance, that is, we just abandon the slowest 11\% of test asteroids in migration. Our calculation shows that this threshold migration speed is $\sim$0.007\,AU/Gyr. Such a simple model gives only a rough estimation and may cause an excessive depletion of asteroids that are large in size (because large asteroids migrate slowly). As a matter of fact, the real transit of asteroids through an MMR could be a complicated process. For example, \citet{mil16} show that below some limiting value of $\dif a/\dif t$, objects often have relatively shorter delay time (arising from the dynamical effect of the MMR) than that of quick transit, although such a phenomenon barely affects the probability of an asteroid being depleted \citep{xu20}. 

\subsection{Space distribution of Eos family members}

With all the parameters determined and the distributions of $\gamma$ and $\omega_{\rm rot}$ set, we calculate the migration of the 14,785 test asteroids from the initial position ($a_0=3.015$\,AU) in $1.3$\,Gyr, so that the final distribution of semimajor axis can be obtained and compared with the real Eos family. 
%because we believe this dispersion will be erased in the subsequent long-term dynamical evolution. 
We remove all the test asteroids that reach the 7/3 MMR and those that pass through the 9/4 MMR with a migration speed slower than $0.007$\,AU/Gyr. Therefore, the final number of test asteroids in our statistics will be smaller than 14,785. 

As mentioned before, we test mainly different distributions of obliquity and rotation rate. For the former, we test constant $\gamma$, random $\gamma$, random $\cos\gamma$, and distribution by \citet{cib16}. While for the latter, we test $\omega_{\rm rot}=b R^{-k}$ for $k= 1/2, 2/3, 1, 3/2, 2$, as well as two cases independent of $R$, i.e. constant $\omega_{\rm rot}$ and the distribution in \citet{warner09}. It should be noted that the coefficient $b$ in each experiment has been very carefully chosen so that the final distribution of $\omega_{\rm rot}$ resembles as much as possible the ``real'' distribution given by \citet{warner09}. 

\begin{figure*}%[htbp]
	\centering
	\includegraphics{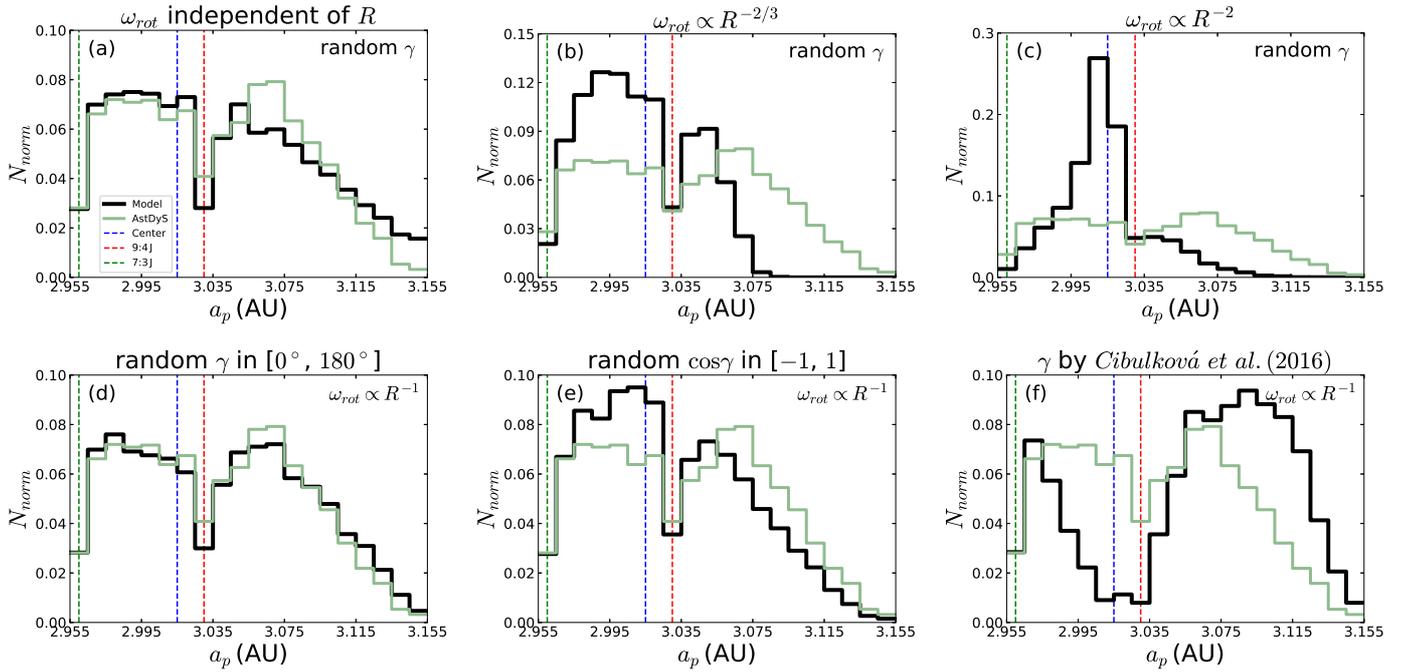}
	\caption{Space distribution of test asteroids mimicking the Eos family (histograms in black). For comparison, the distributions of real family members from {\it AstDyS} are also plotted in cyan. Please note that the ordinate scales in panels (b) and (c) are different from others. In upper panels (a), (b) and (c), the same distribution of random $\gamma\in[0^\circ,180^\circ]$ is adopted but different dependences of $\omega_{\rm rot}$ on $R$ are applied as indicated on the top of each panel. In the lower panels (d), (e) and (f), the same dependence $\omega_{\rm rot}\propto R^{-1}$ is adopted but with different distributions of obliquity as indicated on the top (see text for details). The positions of the family centre, the 9/4 and 7/3 mean motion resonances with Jupiter are indicated by blue, red and green dashed lines. }
	\label{fig:7}
\end{figure*}

Owing to the advantages of the explicit formulae introduced in Section~\ref{sec:est}, the calculation is very easy and fast, making it possible for us to test many combinations of distributions of $\gamma$ and $\omega_{\rm rot}$ with least computation costs. Since the test asteroids are generated according to the real Eos family, their sizes are retrieved by the observed absolute magnitudes $H$, that is, the size distribution is given. Considering meanwhile the thermal parameters, we adopt equation~\eqref{eq:deltarg} to compute the migration $\Delta a$. The final distributions of semimajor axes of the test asteroids are then compared with the real data. We discard most of the results in which the resemblance between the model and reality is poor, and show some typical examples in Fig.~\ref{fig:7}.

Overall, the final semimajor axis distribution of the test asteroids sensitively depends on the distribution of $\gamma$ and the relationship between spin rate $\omega_{\rm rot}$ and size $R$. Apparently, the space distribution in Fig.~\ref{fig:7}(d) matches the real distribution very well. Regarding the assumptions adopted in this panel, i.e. $\omega_{\rm rot}\propto R^{-1}$ and random $\gamma$, as the best resemblance to the reality, we compare all other distributions with this one hereinafter. 

The distribution in Fig.~\ref{fig:7}(a), where random $\omega_{\rm rot}$ is assigned to each test asteroid and all these $\omega_{\rm rot}$ compose a distribution as the one given by \citet{warner09}, also resembles the reality quite well. Compared to Fig.~\ref{fig:7}(d), it has some excess of members both in the region on the left side of the 9/4 MMR and on the far right side, and a deficit can be seen around $a=3.065$\,AU. This implies that in this case both the slow migration and fast migration are more populated than in Fig.~\ref{fig:7}(d). A slow migration leaves more asteroids close to their birthplace (the excess between the 7/3 and 9/4 MMRs), and allows less asteroids to cross the 9/4 MMR (the deficit around $a=3.065$\,AU), while a fast migration sends more asteroids to farther distance (the excess in the far right side). 

In fact, the thermal parameters of Eos family members adopted in our calculations are similar to the ones of regolith-covered objects in Table~\ref{tab:1}, thus we may expect similar dependence of migration distance on $\omega_{\rm rot}$ of Eos members as that of regolith-covered objects. As depicted by the blue line in the middle panel of Fig.~\ref{fig:6}, the migration attains the maximal speed at some specific $\omega_{\rm rot}$ and it drops down in both slower and faster rotation ends. Such nonmonotonic dependence makes it hard to draw a simple conclusion directly from the comparison on which part of the rotation speed distribution has been overestimated. 

Compared to the case $\omega_{\rm rot} \propto R^{-1}$, the rotation rate $\omega_{\rm rot}\propto R^{-2/3}$ indicates that the $\omega_{\rm rot}$ is smaller than in the former case for given small asteroids while it's larger for big ones. Relatively, this produces more slow rotators, thus decreases the migration speed statistically. The distribution in Fig.~\ref{fig:7}(b) for $\omega_{\rm rot}\propto R^{-2/3}$ shows that the test asteroids migrate much slower than in the case of Fig.~\ref{fig:7}(a). Most of them are still gathering in the close vicinity around their birthplace, with the farthest member only reaching $a=3.095$\,AU. Meanwhile, the slow migration causes a big loss of members when passing through the 9/4 MMR.   

For a steeper distribution of rotation rate $\omega_{\rm rot}\propto R^{-2}$, the small asteroids will spin faster while the big ones spin slower, and this shifts the distribution of $\omega_{\rm rot}$ toward larger $\omega_{\rm rot}$ end. As a result, statistically the migration speed decreases even more than in the former case of $\omega_{\rm rot} \propto R^{-2/3}$. As shown in Fig.~\ref{fig:7}(c), a remarkable concentration around the birthplace can be seen, though a small fraction of members may still migrate 0.08\,AU in both directions. 

In the lower three panels in Fig.~\ref{fig:7}, the rotation rate follows the same law $\omega_{\rm rot}\propto R^{-1}$, so that the difference in the space distributions of test asteroids is due to the different obliquity ($\gamma$) distribution. Since the migration speed depends on the obliquity simply by $\cos\gamma$, the comparison among these space distributions may be easily understood. 

The same maximal migration distance corresponding to $\gamma=0^\circ$ as in Fig.~\ref{fig:7}(d) can be found in Fig.~\ref{fig:7}(e), but in this case there are more asteroids of slow migration, implying that the obliquity distribution adopted here (random $\cos\gamma$) underestimates the number of asteroids with $\gamma$ close to $0^\circ$ or $180^\circ$. In fact, the random distributed spin axis orientation (i.e. random $\cos\gamma$) generated by the birth collision will be modified by the YORP effect, which is likely to drive the obliquity $\gamma$ to either $0^\circ$ or $180^\circ$. 

It is still impossible to accurately predict the spin obliquity evolution at present, because the YORP effect is extremely sensitive to the asteroid's shape, of which we unfortunately have only little knowledge. \cite{vokbot12} show that the YORP effect needs $\sim$Gyr to modify the rotation state of asteroids of $\sim 10\,$km in size at a few AU. Therefore, within the age of Eos family, the rotation state of many members must have been modified by YORP effect, and obliquity values shall have accumulated in some extent to either $0^\circ$ or $180^\circ$. Although the collisions may destroy such obliquity accumulation, we still expect to see more obliquities around $0^\circ$ or $180^\circ$ within the family. In this sense, the best fit between the space distributions from our calculation and the reality presented in Fig.~\ref{fig:7}(d), where a random $\gamma$ is adopted, is not just a coincidence. More tests and simulations may put some valuable restrictions on the real distribution of the spin axes orientations of family members, as well as on the collision frequency among them.
 
In Fig.~\ref{fig:7}(f), we show the results calculated from an obliquity distribution given by \citet{cib16}. Apparently, in this case, the migration is too fast. So many members have been driven out from their birthplace, almost leaving a ``gap'' at $a=3.015$\,AU. They migrate so fast that nearly all members successfully cross the 9/4 MMR safely, making an obvious excess of number of members around $a=3.095$\,AU. This means the obliquity distribution in this reference is biased in favour of values close to $0^\circ$ and $180^\circ$. 

To evaluate the similarity between the space distributions calculated from the model and from the observational data, we run the two-sample Kolmogorov-Smirnov (K-S) test implemented in Python\footnote{ https://docs.scipy.org/doc/scipy/reference/generated/scipy.stats.ks\_2samp.html} to obtain the quantitative estimations. For the data summarized in Fig.~\ref{fig:7}(d) that has apparently the best matching, we get the smallest K-S statistic value of 0.0213, implying an identical distribution followed by the two samples. The second best case is the one presented in Fig.~\ref{fig:7}(a) with a K-S statistic of 0.0481, while the other four panels correspond to much higher K-S statistic from 0.136 in panel (e) to 0.382 in panel (c). 

So far, the original dispersion of orbital elements of family members due to the fragmentation that generated the family is ignored in our calculations, which may be an over-simplification. The large parent body of the Eos family implies a large escape velocity, resulting in a considerable orbits dispersion. Some fragments could even be ejected to the right hand side of the 9/4 MMR and a fraction of these objects may migrate inwards, which affects the spatial distribution of family members. 

We do some further calculations to check the influence of initial orbital dispersion. According to \citet{vok06}, we assume the initial velocity dispersion to follow $v\propto R^{-1}$ with $v=93$\,m/s for $R=2.5$\,km. The rotation rate $\omega_{\rm rot}$, obliquity $\gamma$ and other parameters are set in the same way as in Fig.~\ref{fig:7}(d). From such initial orbits we calculate their final positions in term of semimajor axis. The matching between the distribution of these test objects and that of real observational data, with a K-S statistic value of 0.0402, is acceptable but not as good as the one shown in Fig.~\ref{fig:7}(d). Most probably, the result can be greatly improved if the uncertainties in initial orbital dispersion can be reduced and all the involved parameters can be further calibrated. It deserves a thorough investigation in future.

%______________________________________________________________

\section{Conclusions}
The Yarkovsky effect influencing an asteroid's motion depends sensitively on its thermal, physical and dynamical parameters. Adopting the theory proposed by \cite{vok99}, in this paper we analysed the semimajor axis displacement of a celestial body due to Yarkovsky effect.

With appropriate simplification and approximation, we derived the analytical solutions to the perturbation equations for asteroids under the influence of Yarkovsky effect. In these solutions for both seasonal and diurnal effects, the dependences of the semimajor axis displacement ($\Delta a$) on the thermal and dynamical parameters of the asteroid are explicitly given.

The validity and reliability of these solutions are numerically verified by comparison with the direct integration of the equation of motion. The dependence of $\Delta a$ on thermal conductivity $K$, specific heat capacity $C$, bulk density $\rho$, body size $R$, and initial distance to the Sun $a_0$, are carefully discussed. Particularly, we explicitly show that the Yarkovsky drift rate does not decrease monotonically with increasing $a_0$, and the locations of the maximal drift rate in term of semimajor axis are calculated.

The seasonal Yarkovsky effect always drives a celestial body toward the Sun while the direction of the diurnal effect depends on the obliquity of rotation axis. Applying the analytical formula for $\Delta a$ obtained in this paper, we analysed the combined seasonal and diurnal effects and derived the criteria determining the migration direction. The variation of migration distance with respect to obliquity $\gamma$, rotation rate $\omega_{\rm rot}$ and initial semimajor axis $a_0$, are investigated, and the zero points of the migration functions (migration distance as functions of $\gamma, \omega_{\rm rot}$ and $a_0$) suggest that Yarkovsky effect might produce interesting debris distribution profile in the debris disk around a star, unless the YORP effect and mutual collisions break down the balance, which in fact is very likely to happen. 

Owning to the advantages of the analytic formulae, we can easily calculate the migration of asteroids if the thermal and dynamical parameters are given. As an example of the convenient application, we estimate the space distribution of Eos family members. So far, the knowledge about the spinning state of family members, either the rotation rate or the orientation of spin axis, is very limited. Hence the migration history of any individual family member due to Yarkovsky effect is of great uncertainty. But as a whole, the family's space distribution can be statistically obtained if the rotation rate $\omega_{\rm rot}$ and obliquity $\gamma$ follow some reasonable statistical law. Conversely, the comparison between the calculated space distribution from model and the real distribution from observations may help us recognize the real statistical properties of the involved parameters, such as $\omega_{\rm rot}$ and $\gamma$, or at least put some restrictions on them, which in turn can be used to restrict the evolution of the family.

Our calculations suggested that the orientations of spin axes of Eos family members are concentrated to a certain extent in the direction perpendicular to the orbital plane, but not as much as the currently available obliquity data shows, in which the number of members with $\gamma$ close to $0^\circ$ or $180^\circ$ is obviously overestimated. In fact, the obliquities are likely randomly distributed in $[0^\circ, 180^\circ]$, with the orientations of spin axes being just a little biased in favour of being perpendicular to the orbital plane. The originally random orientation of spin axes (random $\cos\gamma$) has been modified later by the competitive processes of collision and YORP effect, hence the current state of random $\gamma$ is achieved. 

Our calculations also suggested that the rotation rates of family members depend on the asteroid size by $\omega_{\rm rot}\propto R^{-1}$. This implies that within the time scale of YORP effect, the collisions among family members are likely to happen frequently.

\section*{Acknowledgements}
We thank the anonymous referee for the helpful comments and suggestions. This work has been supported by the National Key R\&D Program of China (2019YFA0706601) and National Natural Science Foundation of China (NSFC, Grants No.11473016 \& No.11933001).

%-------------------------------------------------------------------

\bsp	% typesetting comment
\label{lastpage}
\end{document}